# Helicity Dependent Directional Surface Plasmon Polariton Excitation Using A Metasurface with Interfacial Phase Discontinuity

**Lingling Huang[1,2*], Xianzhong Chen[1*], Benfeng Bai[2], Qiaofeng Tan[2], Guofan Jin[2], Thomas Zentgraf[3], Shuang Zhang[1]**

1. *School of Physics & Astronomy, University of Birmingham, Birmingham, B15 2TT, UK*
2. *State Key Laboratory of Precision Measurement Technology and Instruments, Department of Precision Instruments, Tsinghua University, Beijing 100084, China*
3. *Department of Physics, University of Paderborn, Warburger Straße 100, D-33098 Paderborn, Germany*

**Corresponding author**:

Shuang Zhang (s.zhang@bham.ac.uk)

School of Physics & Astronomy, University of Birmingham, Birmingham, B15 2TT, UK

**Abstract:**

**Surface plasmon polaritons (SPPs) have been widely exploited in various scientific communities, ranging from physics, chemistry to biology, due to the strong confinement of light to the metal surface. For many applications it is important that the free space photon can be coupled to SPPs in a controllable manner. In this Letter, we apply the concept of interfacial phase discontinuity for circularly polarizations on a metasurface to the design of a novel type of polarization dependent SPP unidirectional excitation at normal incidence. Selective unidirectional excitation of SPPs along opposite directions is experimentally demonstrated at optical frequencies by simply switching the helicity of the incident light. This approach, in conjunction with dynamic polarization modulation techniques, opens gateway towards integrated plasmonic circuits with electrically reconfigurable functionalities.**

**Keywords:** Metamaterials, Surface plasmon, Metasurface

---

[*] These authors contribute equally to this work.

Surface plasmon polaritons (SPPs), the excitation of collective motion of conduction band electrons on metal-dielectric interfaces, have been shown to exhibit intriguing properties of strong enhancement of local field and large in-plane momentum [1, 2]. SPPs can be exploited for versatile functionalities and applications ranging from photonic biochemical sensors [3], non-linear optics [4, 5], solar cells [6, 7], to magneto-optic data storage [8] and sub-diffractional imaging [9]. One particular promising application of SPPs is nanophotonic circuitry with deep subwavelength footprint, benefitting from the combination of high frequency of the light and the strong confinement of the electronic excitation of the electrons. Plasmonic nanophotonic circuits offer opportunities for integrating ultrafast electronic and photonic devices at nanometer scale due to its ability to modify the propagation characteristics below the diffraction limit, such as beaming, focusing, and selective waveguide coupling [10-15].

Coupling of free space photons to SPPs is a key step in plasmonics. Prism coupling, periodic corrugations and topological defects on the surface are the main techniques to realize momentum matching between SPPs and free space photons. Prism coupling, despite being unidirectional, suffers from its large footprint and therefore is not suitable for coupling light to compact integrated plasmonic devices. On the other hand, excitation of SPPs by singly-periodic gratings and topological defects are in general symmetric, leading to two SPPs propagating along opposite directions. Asymmetrical excitation of SPPs has evoked enormous interest recently. Unidirectional SPPs have been achieved by asymmetric design of the grating coupler [16, 17], highly compact plasmonic antennas [18-20], and metasurfaces with locally controlled phase profile [21]. However, the direction of SPP excitation in these couplers is predefined and cannot be reconfigured. Recently, a number of techniques have emerged for tunable unidirectional launching of SPPs. These include using oblique incidence onto slits [22], or utilizing delicate control of phase interaction through blazed gratings, Bragg reflection, and coherence processes [23, 24]. The implementation of these techniques requires either complex optical setups to modulate the retardation between two light pulses, or mechanical adjustments. Here we propose and realize a novel scheme of tunable unidirectional excitation of surface plasmon polaritons whose propagating direction depends on the helicity of incident light, based on a polarization dependent phase discontinuity resulting from carefully designed metasurfaces. This method, in conjunction with dynamic polarization modulation techniques, paves the way towards electrically reconfigurable plasmonic circuits.

Innovative applications of plasmonic meta-surfaces with abrupt phase discontinuities have received much attention recently [25-28]. Multiple functionalities have been experimentally achieved at visible and infrared wavelength ranges, covering generalized refraction [25, 26], vortex beam generation [25], flat meta-lenses [27, 28], and broadband quarter wave plates [29]. In particular, circular polarization (helicity) based metasurfaces have shown intriguing dispersionless broadband properties as well as polarization dependent refractions, through careful arrangement of dipole antennas with spatially varying orientations [30]. Here we employ a plasmonic metasurface to realize helicity dependent SPP unidirectional excitation. By changing the helicity of incident beam, we can achieve an asymmetric momentum matching condition of SPPs along different directions for input light at normal incidence.

The unidirectional SPP coupler consists of an array of elongated apertures with a constant gradient of orientation angle $\varphi$ along $x$ direction, as shown in Fig. 1(a). In general, each aperture can be considered as a combination of electric and magnetic dipoles. Each dipole, under the illumination of circularly polarized light at normal incidence, can be decomposed into two circularly oscillating components, with one component having the same helicity as the incident light and a phase that does not depend on the orientation of the aperture, and the other with opposite helicity to the incident light and a phase $2\varphi$ that is twice the orientation angle of the aperture (see Supplementary Materials). For an array of apertures with a constant gradient in the orientation angle, the refraction and diffraction of light by the array show ordinary and anomalous refraction and diffraction orders [30]. In particular, such phase discontinuity for anomalous refraction and diffraction is geometric in nature and does not rely on the incident wavelength. By taking into account the contribution from the phase gradient, the angles of the ordinary and anomalous refracted and diffracted beams for a normal incident beam are given by [31]:

$$n_t \sin\theta_t = m\frac{\lambda_0}{s} \qquad \text{(Ordinary diffraction)} \qquad (1a)$$

$$n_t \sin\theta_t = m\frac{\lambda_0}{s} + 2\sigma\frac{\lambda_0}{2\pi}\frac{\Delta\varphi}{s} = \frac{\lambda_0}{s}(m + 2\sigma\frac{\Delta\varphi}{2\pi}) \qquad \text{(Anomalous diffraction)}$$

(1b)

where $n_t$ and $\theta_t$ are the refractive indices and angles for the transmitted (index 't') beams, respectively. $\lambda_0$ is the incident wavelength in vacuum, $m$ is the diffraction order, and $\sigma$ represents the helicity of the incident beam, which takes the values of ±1. Eqn (1a) and Eqn (1b) describe the ordinary and anomalous diffraction, respectively. Note that for anomalous refraction and diffraction, the sign of the phase change depends on the polarization state of the incident beam.

Fig. 1(b) illustrates the ordinary and anomalous refraction and diffraction for the two circular polarization states. For a circularly polarized incident beam at normal incidence, the two ordinary first-order diffracted beams are symmetric about the central ordinary zero order beam along the surface normal. In contrast, the anomalous refracted beam and two first order anomalous diffracted beams are shifted towards the same direction relative to their ordinary counterparts. As a result, the two anomalous first-order diffracted beams are not symmetric about the surface normal, which forms the basis of unidirectional SPP excitation at certain optical frequencies.

The lattice constant between neighboring apertures $s$ and the step of the rotation angle of the aperture $\Delta\varphi$ offer the necessary phase matching condition for a beam at normal incidence to excite SPPs, as illustrated by Figure 1(c). The in-plane wave vector of light for generating SPPs can be calculated by:

$$k_{spp} = \frac{2m\pi}{s} + \sigma \cdot \frac{2\Delta\varphi}{s} \qquad (2)$$

For ordinary refraction, both of the two first-order diffracted beams ($m = \pm1$) couple to SPPs at the same optical frequency $\omega_2$ ($\lambda_2$), but with opposite propagation directions. For anomalous refraction with an incident beam of right handed helicity ($\sigma = +1$), the phase matching condition is shifted to a higher frequency $\omega_1$ ($\lambda_1$) for SPPs propagating along $+x$ direction, and to a lower frequency $\omega_3$ ($\lambda_3$) for propagating along $-x$ direction. This effect arises due to the extra positive in-plane momentum arising from the phase gradient. Interestingly, when the helicity of the input beam is changed to left handedness ($\sigma = -1$), the phase matching condition is reversed, leading to unidirectional SPP propagation along $+x$ and $-x$ at $\omega_3$ and $\omega_1$, respectively. Thus, at the two optical frequencies $\omega_1$ and $\omega_3$, the propagation direction of SPPs can be reversed by simply changing the circular polarization of the input light, as illustrated by Fig. 1(d).

For a metal film on a glass substrate, SPPs are supported at both the air/metal and glass/metal interfaces. In general, there is a coupling between the SPPs at the two interfaces to form hybrid modes. However, as the metal film is 40 nm thick, which is twice the skin depth, the coupling between the two interfaces is very small, leading to decoupled SPPs at the top and bottom interface. Here we only investigate the excitation of SPPs at the glass/metal interface. The dispersion curve of SPPs at the glass/metal interface is given by:

$$\lambda_{spp} = \lambda_0 \sqrt{\frac{\text{Re}(\varepsilon_m) + \varepsilon_d}{\text{Re}(\varepsilon_m) \cdot \varepsilon_d}} \qquad (3)$$

where $\lambda_0$ is the free space wavelength of incident light, $\varepsilon_m$ and $\varepsilon_d$ are the permittivities of the metal and the glass substrate, respectively. In the following we will use gold as the metal whose permittivity is described by the Drude model, with a plasmon frequency $\omega_p$ = 1.366 $\times 10^{16}$ rad/s, and damping frequency $\gamma$ = 1.2$\times 10^{14}$ rad/s. For the design of the metasurface, we set the lattice constant $s$ to be 570 nm, and the step-size between the orientations of neighboring apertures to be $\pi/6$. This leads to a periodicity of the metasurface of 6 × 570nm = 3.42 μm. For this configuration the wavelengths at which SPPs can be excited for normal incident light can be calculated by using Eqn.2 and Eqn.3 (dispersion curve shown in Fig.2): $\lambda_2$=860 nm corresponds to a symmetric excitation of counter propagating SPPs, and $\lambda_1$=762 nm and $\lambda_3$=1016 nm correspond to the polarization dependent unidirectional excitations.

In the following, we experimentally prove our theoretical design by demonstrating the helicity dependent unidirectional coupling of light to SPPs. The sample was fabricated on a quartz glass substrate (with refractive index n=1.45) with standard electron-beam lithography, followed by metal deposition (3 nm Ti and 40 nm Au) and a lift-off process. A scanning electron microscopy (SEM) image of the fabricated sample is shown in Fig. 3(a). The nano-apertures with variable orientation are ~200 nm long and ~50 nm wide, and the lattice constant $s$ = 570 nm is the same along both $x$ and $y$ directions. The nano-aperture array covers an area of $17\times100$ μm$^2$, containing 5 periods along the direction of SPP excitation. In order to scatter the SPPs into the far-field for simple detection, two gratings were symmetrically located on both sides of the array as out-couplers. The gratings have a pitch of 400 nm with duty cycle of 50% and a length of 100 μm. Each out-coupler is 25 μm away from the edge of the nano-aperture array.

We employ far-field microscopy detection to experimentally demonstrate the directional excitation of SPPs at the glass/metal interface. The experimental setup is schematically illustrated in Fig.3 (b). The CP light with left or right circular polarizations, generated by a polarizer and a quarter-wave plate, is normally incident on the front side (air/metal interface) of the sample. Two objectives are confocally aligned for the excitation and the detection of the scattered SPP wave. The scattered light emanating from the out-coupler gratings is collected with a $20\times/0.40$ objective and imaged to a CCD camera.

Figure 4 (a-e) show the obtained far-field images of the scattered light from the SPPs by the output-couplers for circularly polarized incident light at wavelength of 1020 nm, 870 nm and 780 nm, respectively. Note that the wavelengths where SPPs are observed are slightly shifted away from the designed values because of the possible deviation of geometries of the fabricated sample from the design. The excitation of the SPPs at these three wavelengths is evidenced by the elongated bright spots located at the positions of the out-couplers. Away from these wavelengths, the phase matching condition is not exactly matched, and the conversion efficiency to SPP drops correspondingly. The white saturated spots at the center correspond to the directly transmitted light through the nano-apertures and the metal film. As shown in Fig. 4(a) and (b) for $\lambda$ = 1020 nm, the propagation direction of the SPPs can be reversed to the opposite directions when the circular polarization of the incident beam is switched. While in Fig. 4(c) for $\lambda$ = 870 nm, the SPPs are excited equally along both directions, due to the fact that the momentum conservation for both directions are matched simultaneously for the two first-order ordinary refracted beams. The switchable SPP excitation is also observed for $\lambda$=780 nm, despite that the contrast between SPPs propagating to the opposite directions is not as high as that for $\lambda$ =1020 nm. We plot the line cross sections of the intensity corresponding to each case as well, where the helicity dependent unidirectional excitation at anomalous diffraction orders are further verified. All these results are in good agreement with full wave simulations calculated by COMSOL Multiphysics shown in Fig. 4(f-j). Due to the presence of a significant level of background optical intensity, it is difficult to accurately characterize experimentally the extinction ratio between the coupling efficiency of SPPs along the two opposite directions. Numerical simulations show that at $\lambda_1$=762 nm the extinction ratio is around 24, and at $\lambda_3$=1016 nm the extinction ratio can reach 270. It is also numerically calculated that the excitation efficiency for such a nano-aperture array is around 3.8% (see Supplementary Materials). This low efficiency is mainly because that the structure we employed (holes in metal film) is not an optimized configuration, but it is chosen because of the ease of fabrication. It is expected that the coupling efficiency would be significantly enhanced by replacing the non-resonant apertures with resonant structures, such as the magnetic antennas exhibiting a strong magnetic resonance as demonstrated by Ref. 19. Theoretically we have investigated a metasurface consisting of metal/dielectric/metal resonators, which give rise to significantly higher efficiency of 14% (see Supplementary Materials).

As any polarization state can be decomposed into left and right handed circular polarizations, it is expected that arbitrary ratios between the excitation of SPPs propagating along opposite directions can be achieved by simply adjusting the ellipticity of the polarization state for the incident beam. This is experimentally confirmed, with the results shown in Fig. 5. By rotating the axis of the quarter-wave-plate while keeping the orientation of the linear polarizer fixed at 0°, the polarization state can be tuned continuously from one circular polarization, through elliptical and linear polarizations, to the opposite circular polarization. The ellipticity of light is defined as $\eta = (E_R - E_L)/(E_R + E_L)$, with ±1 representing right (left) circular polarization, and 0 representing linear polarization. The ellipticity $\eta$ is related to the rotation angle of the axis of the quarter wave plate $\theta$ simply by $\eta = \tan(\theta)$. Fig. 5 shows continuously tunable SPP excitation efficiencies between the two opposite directions at $\lambda$ = 1020 nm. For left/right circular handedness at points A and G, the SPPs are propagating at a predominant direction, while for ellipticity equal to zero (point D), equal excitation along both directions are achieved. The experimental results are in good agreement with the numerical simulation (solid curves) that is offset by the background optical intensity. Such ellipticity tunable properties can great facilitate transforming desired distribution of input energies into a plamonic circuits.

In summary, we have demonstrated a polarization dependent unidirectional SPP excitation due to phase discontinuities introduced by an array of plasmonic apertures with spatially varying orientations. The directional SPP excitation shows very high extinction for circularly polarized incident light. The ratio for the excitation of SPPs propagating to opposite directions can be simply adjusted by the ellipticity of the input light. In comparsion to a previous work that uses metasurface to launch unidirectional SPP with high efficiency [21], our work focuses on a totally different functionality – spin switching of the direction of the surface plasmon excitation. The physics underlying our work is a Berry geometrical phase that depends solely on the orientation of each nano-aperture, but not the structure and shape of each individual element. Importantly, as the switching of the SPP excitation is based on the polarization of input light, polarization modulator based on liquid crystals or polymers [31] can be monolithically incorporated to form a compact, electrically controlled plasmonic circuit. The device works upon normal incidence, which would facilitate realistic experimental setup for the next generation of integrated plasmonic circuits.

**Figure Captions**:

**Figure. 1** (a) Schematic of a unidirectional SPPs coupler. The coupler consists of an array of rectangular apertures with spatially varying orientations on a metal film. (b) Ordinary and anomalous refraction and diffraction for the two circular polarization states. The anomalous diffraction orders are asymmetric about the surface normal. (c) Dispersion curve of SPPs and the momentum matching condition for ordinary and anomalous diffraction orders. (d) Only one of the anomalous diffraction orders in (b) can be matched to the SPP dispersion relation to launch unidirectional SPP due to the asymmetry in the anomalous diffraction orders. Interestingly, when the helicity of the incident beam is reversed, so is the direction of the SPP excitation.

**Figure. 2** Dispersion curves for SPPs at air/metal and metal/glass interfaces, respectively. The excitation frequencies are plotted in coincident with the ordinary and anomalous diffraction orders. Red and blue star symbols represent the phase matching condition due to the anomalous diffraction for RCP and LCP, respectively. Green star symbols, which are symmetrically located, corresponds to +1 and -1 ordinary diffraction orders.

**Figure. 3** SEM image of sample and scheme of experimental setup. (a) An SEM image of the gold nano-aperture array with slit gratings on both sides; the right part image is with higher magnification to show the detail; Each gold nano aperture is ~200 nm long and ~50 nm wide, and the lattice constant $s$ between two slits along the $x$ and the $y$ direction is 570 nm; the nano aperture array covers an area of $17 \times 100$ $\mu m^2$. The symmetrically located slits as out-couplers have a pitch of 400 nm with duty cycle of ½ and a length of 100 $\mu m$. Each out-coupler is 25 $\mu m$ away from the edge of the dipole array. (b) far-field imaging detection with microscopy setup. The light acquires a circular polarization after passing through a linear polarizer is along the vertical (+y), and a quarter wave plate with axis along ±45°. The beam is then focused onto the array of nano-apertures by an objective. The surface plasmons, scattered by the output grating coupler, are collected another objective, and imaged onto a CCD camera.

**Figure.4** Experiment and simulation demonstration of selective unidirectional SPP generation (a, f) Experimental observation at $\lambda$ =1020 nm and simulation at $\lambda_3$ =1016 nm for an incident beam with right handed circular polarization (RCP). SPPs are only excited along +$x$ direction. (b, g) The same as (a, f), but for an incident beam with left handed circular polarization (LCP). The SPP is only excited along −$x$ direction. (c, h) Experimental observation at $\lambda$ =870 nm and simulation at $\lambda_2$ =860 nm, where SPPs are excited along both directions. Note that the results are the same for both left and right-handed circular polarizations. (d, i), (e, j) Experimental observation at $\lambda$ =780 nm and simulation at $\lambda_1$ =760 nm, with opposite helicities. Line cross sections are plotted with correspondence to each experiment demonstration.

**Figure. 5** Experiment demonstration of continuously tunable contrast between SPP excitation along +x and −x directions by varying the ellipticity of the input light at $\lambda$ =1020 nm. Points A-G correspond to the CCD images for $\eta$=−1 ($\theta$=−45°), $\eta$=−0.364 ($\theta$=−20°), $\eta$=−0.176 ($\theta$=−10°), $\eta$=0 ($\theta$=0°), $\eta$=0.176 ($\theta$=10°), $\eta$=0.364 ($\theta$=20°), $\eta$=1 ($\theta$=45°). The symbols show the y-integrated output power at the two positions for out-coupler vs. ellipticity ranging from [−1, 1]. The solid curves are calculated by using full wave simulation by COMSOL Multiphysics to investigate the peak intensities at the metal and dielectric boundaries at each side. All intensities are normalized in the range of [0, 1].

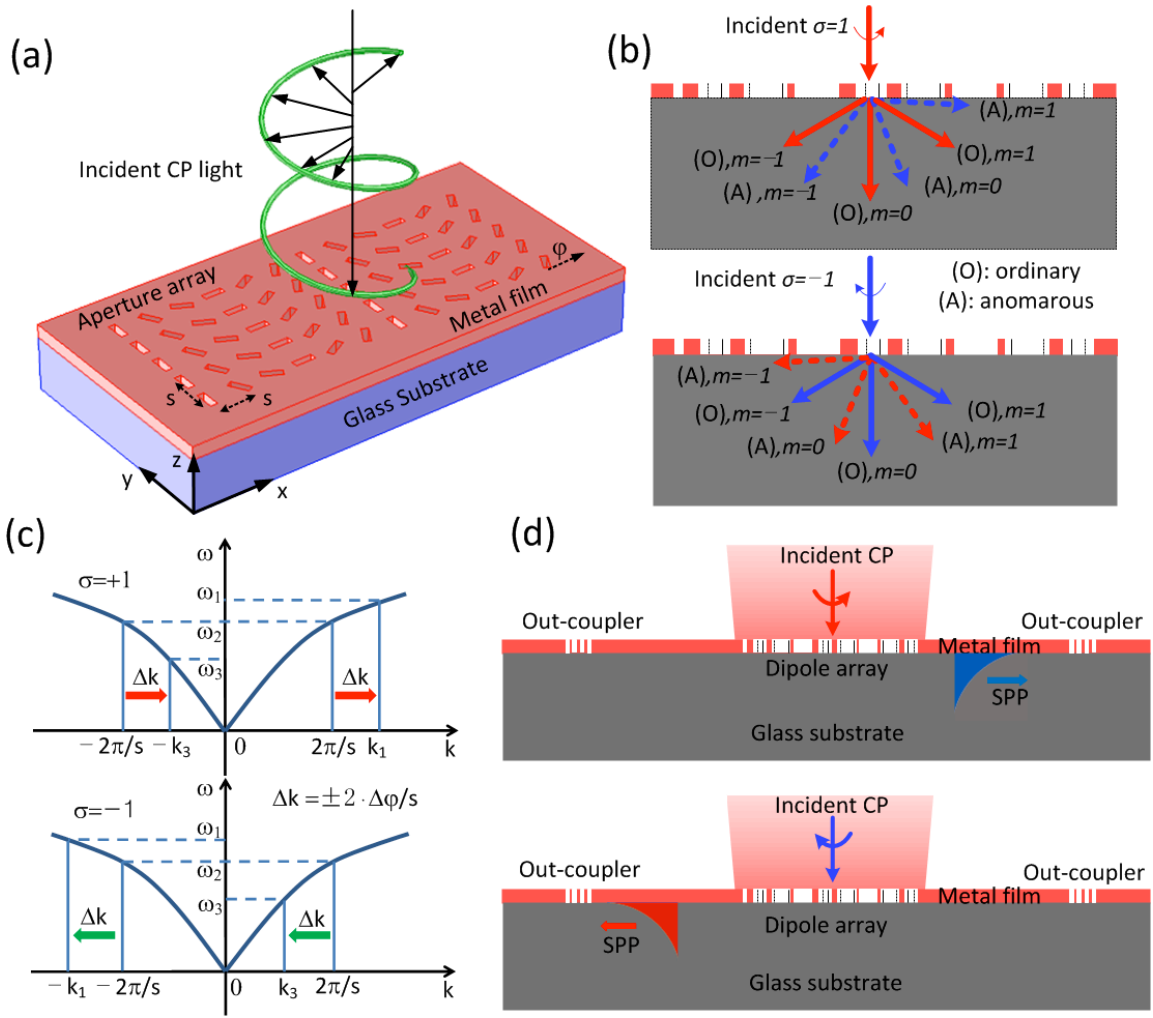

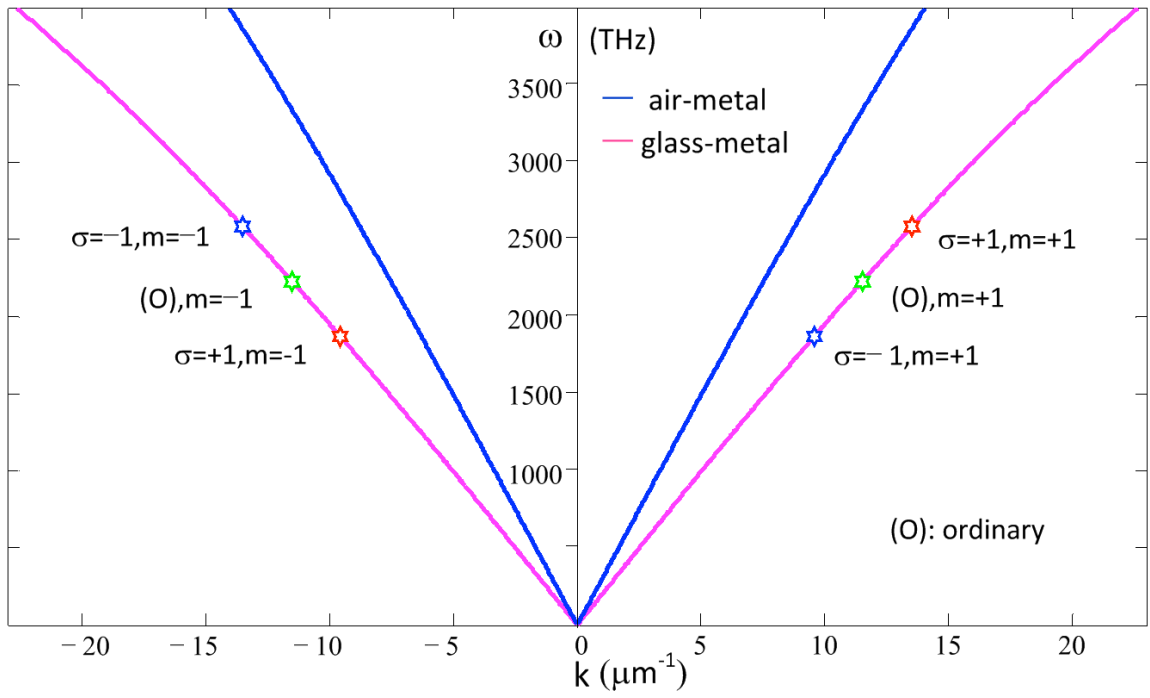

(a)

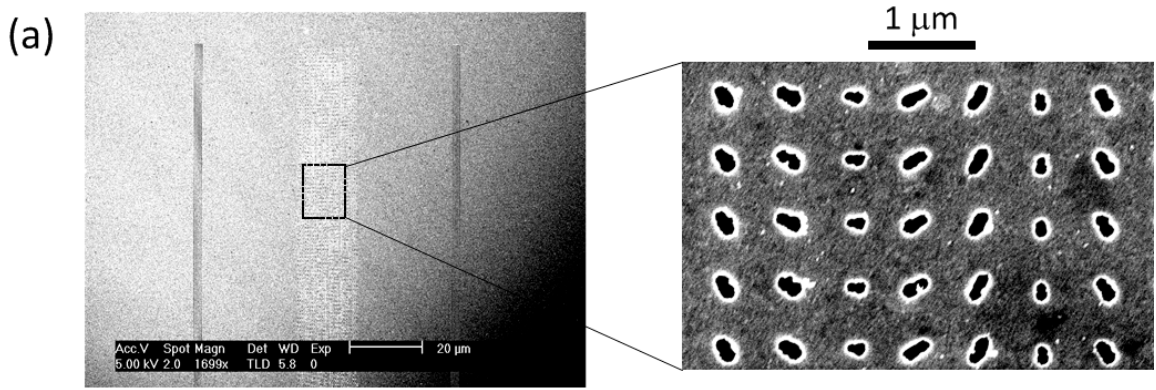

(b)

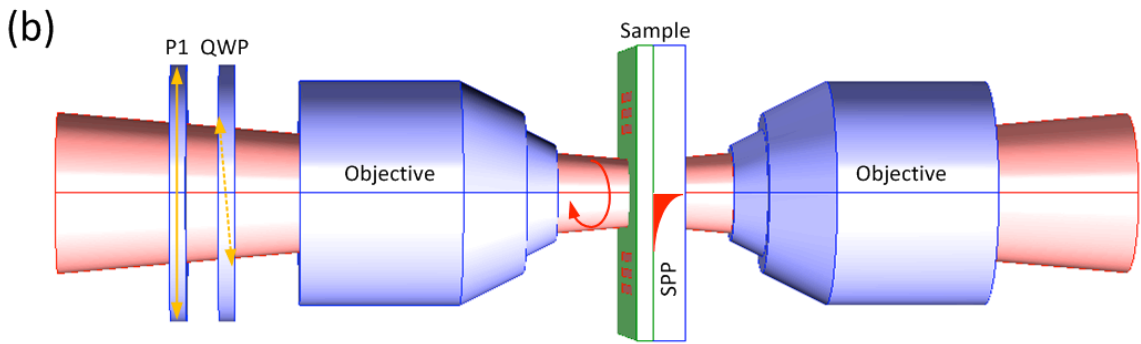

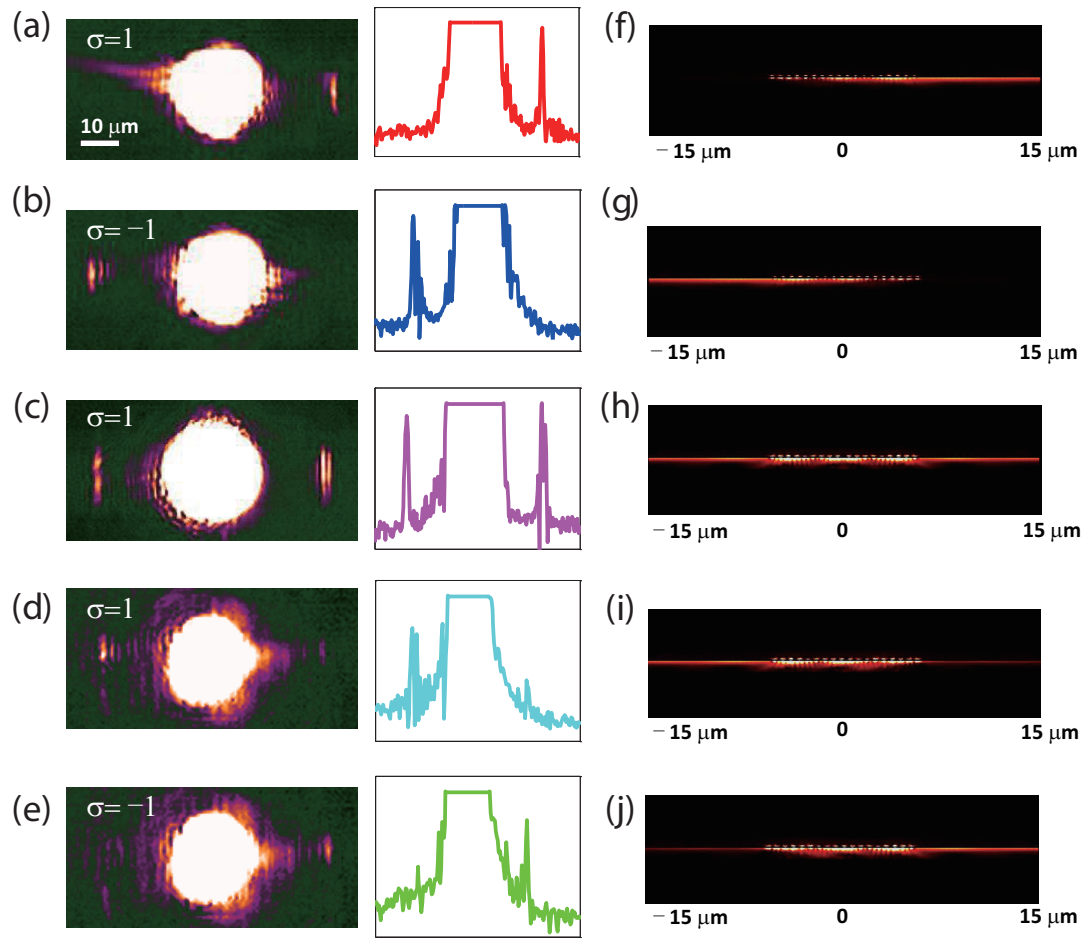

A

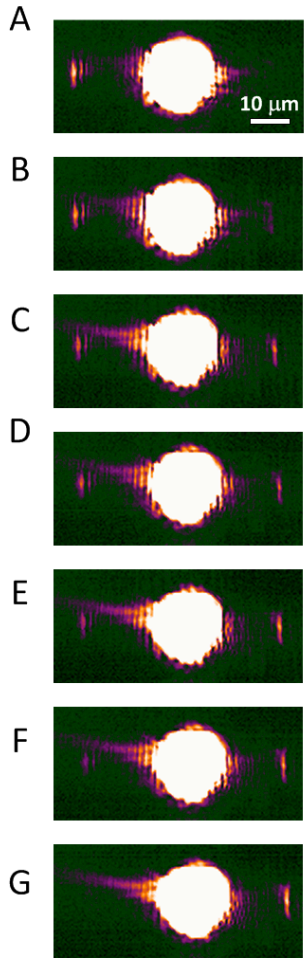

B

C

D

E

F

G

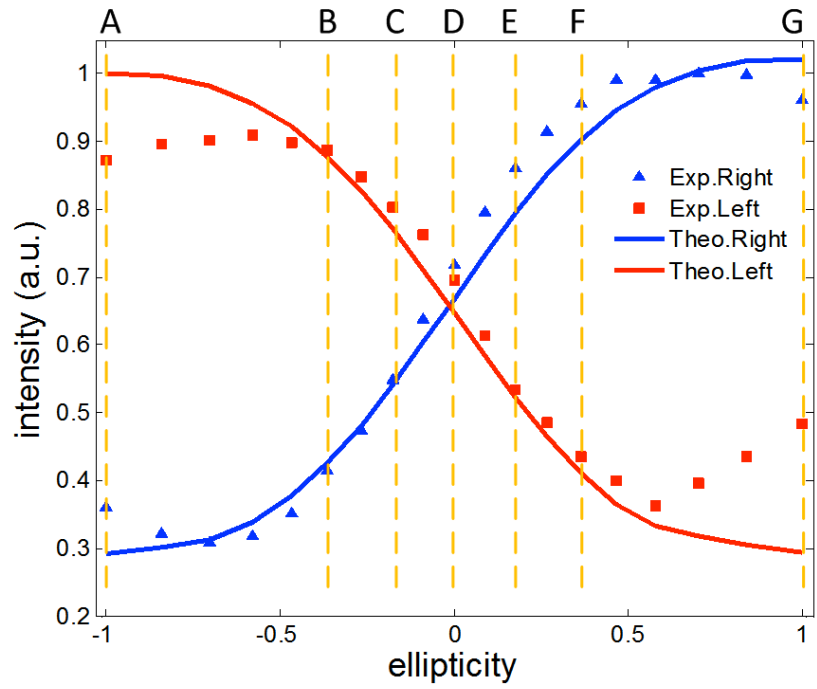